\documentclass[aps,prl,twocolumn,superscriptaddress,longbibliography,floatfix]{revtex4-2}
\usepackage{amsmath,amssymb,bm,graphicx,xcolor}
\usepackage{placeins}
\usepackage[colorlinks=true,citecolor=blue,urlcolor=blue,linkcolor=blue]{hyperref}

\AtBeginDocument{\raggedbottom}

\newcommand{\Lcal}{\mathcal L}

\begin{document}
\title{Spin conservation protects the altermagnetic spin-splitter current from critical scattering}
\author{Yunxi Jiang}
\affiliation{Universit\'e de Strasbourg, CNRS, Institut de Physique et Chimie des
Mat\'eriaux de Strasbourg, UMR 7504, F-67000 Strasbourg, France}
\author{Hao Yu}
\email{Hao.Yu@xjtlu.edu.cn}
\affiliation{Department of Physics, Xi'an Jiaotong-Liverpool University, 215123, Suzhou, China}
\begin{abstract}
Near a magnetic transition, critical fluctuations scatter electrons strongly, but the charge current survives, because long wavelength fluctuations barely turn the electron velocity. In an altermagnet the electric field also drives a spin current, the spin splitter effect, and a nearly forward scattering event that flips spin moves an electron to the other spin band at almost unchanged velocity. We show that the spin current shares the protection of the charge current only when the fluctuations conserve spin. Solving the Boltzmann equation with the full collision operator, including its change with the magnetic order, we find that forward flips equalize the two spin populations at equal energy at the quasiparticle rate, which in two dimensions grows linearly with the correlation length, whereas spin conserving fluctuations relax the spin current at a rate comparable to the charge relaxation rate. Fluctuations with the same resistivity can thus move the spin splitter ratio in opposite directions. Energy conservation protects the part of the spin current that follows the band shift of the occupied states, and the splitting blocks forward flips once the spin Fermi lines lie farther apart than the inverse correlation length. A disorder series at fixed reduced temperature measures the initial fall of the ratio with the magnetic resistivity independent of the order amplitude, and the growth of this slope toward the transition identifies spin flip critical fluctuations.
\end{abstract}
\maketitle

In an altermagnet, collinear order with zero net magnetization splits the bands by spin with a momentum dependence of $d$-wave or higher symmetry \cite{PhysRevX.12.040501,Naka2019,PhysRevLett.126.127701}, so an electric field drives a spin current along with the charge current, the spin splitter effect. Spin splitting torques in metallic bilayers have been studied \cite{Bai2022,Bose2022,Karube2022,Bai2023}, although the magnetic state of RuO$_2$ and the interpretation of its transport signals remain debated \cite{Hiraishi2024,Kessler2024,Liu2024Ru,Wang2026Ru}. The spin splitter ratio $\theta$ of spin to charge conductivity is usually computed with a constant relaxation time \cite{Dou2025}, which multiplies both conductivities by the same factor and so fixes $\theta$ from the bands alone.
But scattering can change spin responses qualitatively. Impurity vertex corrections cancel the intrinsic spin Hall current of the Rashba gas \cite{Inoue2004}, and phonons can enhance, suppress or reverse the spin splitter effect \cite{He2025,Hodt2026}.

Near the ordering temperature $T_N$, the strongest scatterers are the fluctuations of the order itself. They broaden the electronic states, but the charge current survives, because their long wavelength part scatters electrons mostly through small angles \cite{Fisher1968,DellAnnaMetzner2006,DellAnnaMetzner2007,WangBerg2019}. 
This raises the question of whether the spin current survives as well. 
The two spin bands of an altermagnet have equal densities of states but different Fermi lines, so a nearly forward spin flip moves an electron to the other band at almost unchanged velocity. 
In ferromagnets, spin flips mix the two spin channels and change the resistivity only when they conduct differently \cite{FertCampbell1968,FertCampbell1976}. 
In an altermagnet the channels differ only at linear order in the splitting, so at the order at which the spin current appears, spin flip and spin conserving fluctuations of equal total strength give the same resistivity.

We have solved the Boltzmann equation with the full collision operator of quasistatic magnetic fluctuations, including its change with the magnetic order, and compared it with lifetime and transport time approximations built from the same scattering \cite{Ponce2020,Claes2022}.
We find that the protection depends on the spin of the fluctuations.
Spin conserving fluctuations relax the spin current at a rate comparable to the charge relaxation rate, whereas nearly forward flips equalize the two spin populations at equal energy at the quasiparticle rate. 
In two dimensions,
this rate grows linearly with the correlation length $\xi$. Fluctuations with the same resistivity can move $\theta$ in opposite directions,
and the transport time approximation, which reproduces the initial growth of the resistivity, recovers only a quarter to a half of the initial change of $\theta$.
Two effects limit the loss.
Energy conservation on the split bands protects the part of the spin current that follows the band shift of the occupied states, and the splitting blocks forward flips once the two spin Fermi lines lie farther apart than $1/\xi$, which eventually happens on approaching $T_N$ because $\xi$ diverges faster than the splitting vanishes.
How fast $\theta$ falls with the magnetic resistivity, and how this rate grows toward $T_N$, reveals whether the critical fluctuations flip spin.

\emph{Model.}
For spin $\sigma=\pm1$ along a fixed axis and a dimensionless amplitude $\phi_0$ of the magnetic order, the two bands are
\begin{align}
\varepsilon_{\mathbf k\sigma}&=\epsilon_0(\mathbf k)-\mu+\sigma\phi_0d_{\mathbf k},\nonumber\\
v_i&=\hbar^{-1}\partial_{k_i}\epsilon_0,\quad u_i=\hbar^{-1}\partial_{k_i}d_{\mathbf k}.
\label{eq:l_band}
\end{align}
Here $v_i$ is the velocity of the unsplit band and $u_i$ its change per unit $\sigma\phi_0$. On the square lattice, $\epsilon_0=-2t_1(\cos k_x+\cos k_y)-4t_2\cos k_x\cos k_y$, with nearest neighbor hopping $t_1$ and next nearest neighbor hopping $t_2$, and $d_{\mathbf k}=d_0(\cos k_x-\cos k_y)$ with exchange scale $d_0=t_1$. 
The reference bands are the band with only nearest neighbor hopping (NN band), $(t_2/t_1,\mu_0/t_1)=(0,-1)$, and the band that also includes next nearest neighbor hopping (NNN band), $(-0.2,-1.2)$, where the zero temperature chemical potential $\mu_0$ fixes the density. 
Both use $k_BT/t_1=0.05$ and $\xi\bar k_F=4$ unless $\xi$ is varied, with $\bar k_F$ the mean Fermi radius and $\xi$ the magnetic correlation length, which sets the typical momentum transfer $1/\xi$. 
The parabolic reference band has $\epsilon_0=\hbar^2k^2/(2m)$ and $d_{\mathbf k}=d_0(k_x^2-k_y^2)/k_F^2$, with $m=d_0=\mu_0=\hbar=1$ and $k_BT/\mu_0=0.05$.

Longitudinal and transverse magnetic fluctuations are uncorrelated. 
The Born collision operator then keeps the distribution diagonal in spin, and spin flips couple the occupations of the two spins. A chiral transverse correlator makes opposite flips unequal only through inelastic dynamics, so its contribution is odd in the order and suppressed by $\hbar\omega_q/k_BT$ in the elastic quasistatic limit, as shown in the End Matter.

We write the occupation driven by the field as $\delta f_\sigma=-eE_x\tau_{\rm bg}(h_++\sigma\phi_0h_-)$, where $\tau_{\rm bg}$ is the scattering time of disorder and other nonmagnetic processes, $F=-f'$ and $F'=dF/dE$ at $E=\epsilon_0-\mu$. Differentiating the Boltzmann equation with respect to $\phi_0$ gives
\begin{align}
h_+&=(I+\tau_{\rm bg}\Lcal_+)^{-1}X_0,\nonumber\\
h_-&=(I+\tau_{\rm bg}\Lcal_-)^{-1}(X_1-\tau_{\rm bg}\Lcal_1h_+).
\label{eq:l_kinetic}
\end{align}
And $X_0=Fv_x$ and $X_1=Fu_x+d_{\mathbf k}F'v_x$. 
The operators $\Lcal_\pm$ relax the parts of the distribution common to the two spins and opposite between them, and the background contributes the identity. 
$X_1$ contains the change of velocity and the energy shift of the occupied states. 
The last term, $-\tau_{\rm bg}\Lcal_1h_+$, arises because the collisions change with the order. 
An elastic transition $(\mathbf k,\sigma)\to(\mathbf k',\sigma')$ reaches final states whose unsplit energy is shifted by $\sigma\phi_0(d_{\mathbf k}-\sigma\sigma'd_{\mathbf k'})$. 
A nearly forward transition that conserves spin carries the weight $d_{\mathbf k}-d_{\mathbf k'}\simeq0$, whereas a nearly forward flip carries $d_{\mathbf k}+d_{\mathbf k'}\simeq2d_{\mathbf k}$ and lands on the other spin band.

With the spin current in electrical units, $J^s=-(2e/\hbar)J^{S_z}$, the spin and charge conductivities are $\sigma^s_{xx}=2e^2\tau_{\rm bg}\phi_0S$ and $\sigma^c_{xx}=2e^2\tau_{\rm bg}C$, with
\begin{equation}
S=\int_{\mathbf k}(u_xh_++v_xh_-),\quad C=\int_{\mathbf k}v_xh_+,
\label{eq:l_current}
\end{equation}
where the measure counts one state per spin and unit area. Their ratio is $\theta=\phi_0Q$, where $Q=S/C$ is the reduced spin splitter ratio with the order amplitude removed. With background scattering alone, $S$ and $C$ take the values $S_0$ and $C_0$, and $Q$ reduces to $Q_{\rm CRT}=S_0/C_0$.

Longitudinal fluctuations, which conserve spin, and transverse fluctuations, which flip it, scatter electrons with the same momentum dependence $|\Lambda_m|^2/(\xi^{-2}+q^2)$ \cite{DellAnnaMetzner2006,DellAnnaMetzner2007,WangBerg2019} and with strengths $A_L$ and $A_T$. The coupling $\Lambda_A=(\Phi_d+\Phi_d')/2$, with $\Phi_d=d_{\mathbf k}/d_0$, averages the form factor over the initial and final momenta, and the factor $e^{-q^2/8}$ in $\Lambda_G=\Lambda_Ae^{-q^2/8}$ suppresses scattering with large momentum transfer. 
The ratio $R=A_T/A_L$ sets the fraction $R/(1+R)$ of scattering events that flip spin. 
Equal fluctuations of the three spin components give $R=2$, critical longitudinal fluctuations under easy axis anisotropy give small $R$, and soft transverse modes in the ordered phase give large $R$. 
Both kinds of fluctuations share the correlation length $\xi$, and Ref.~\cite{JiangCompanion2026} treats unequal lengths. The kinetic and quasistatic descriptions require $\hbar\Gamma_{\rm out}\ll E_F$, $|\mathbf v|/\Gamma_{\rm out}>\xi$ and $\hbar\omega_q\ll k_BT$. The End Matter discusses these conditions and defines the dimensionless spin conserving and spin flip strengths $K$ and $a$.

\emph{Spin and charge current decay.}
To first order in the magnetic rates, $C\simeq C_0(1-\tau_{\rm bg}\Gamma_c)$ and $S\simeq S_0(1-\tau_{\rm bg}\Gamma_s)$, with
\begin{align}
\Gamma_c&=\langle v_x,\Lcal_+X_0\rangle/C_0,\nonumber\\
\Gamma_s&=[\langle u_x,\Lcal_+X_0\rangle+\langle v_x,\Lcal_-X_1\rangle+\langle v_x,\Lcal_1X_0\rangle]/S_0.
\label{eq:l_rates}
\end{align}
The ratio then changes as
\begin{align}
1-Q/Q_{\rm CRT}&=s\rho_{\rm mag}/\rho_{\rm bg}+O(\rho_{\rm mag}^2/\rho_{\rm bg}^2),\nonumber\\
s&=\Gamma_s/\Gamma_c-1.
\label{eq:l_slope}
\end{align}
The slope $s$, the fractional drop of $\theta$ per unit relative magnetic resistivity $x=\rho_{\rm mag}/\rho_{\rm bg}$, compares the rates at which the spin and charge currents start to decay, and positive $s$ means that the spin current decays faster. The term with $\Lcal_1$ enters only $\Gamma_s$, because the charge conductivity is even in $\phi_0$.

At zero splitting, we write $D$ for scattering out of a state and $G$ for scattering into it, common to both kinds of fluctuations. A flip exchanges the spin labels, so transverse scattering into a state enters the spin difference with the opposite sign,
\begin{align}
\Lcal_+&=(A_L+A_T)(D-G),\nonumber\\
\Lcal_-&=A_L(D-G)+A_T(D+G).
\label{eq:l_operators}
\end{align}
Magnetic scattering with a given total strength $A_L+A_T$ gives the same charge conductivity for every $R$, whereas $\Lcal_-$ and $\Lcal_1$ depend on $R$. Because $\Gamma_c$ and $\Gamma_s$ are linear in the strengths,
\begin{equation}
s(R)=\frac{s_L+Rs_T}{1+R},
\label{eq:l_channels}
\end{equation}
where $s_L$ and $s_T$ are the slopes for purely spin conserving and purely spin flip scattering. As the fraction of flips grows, $s$ moves from $s_L$ to $s_T$.

\begin{figure*}[tbp]
\centering\includegraphics[width=\textwidth]{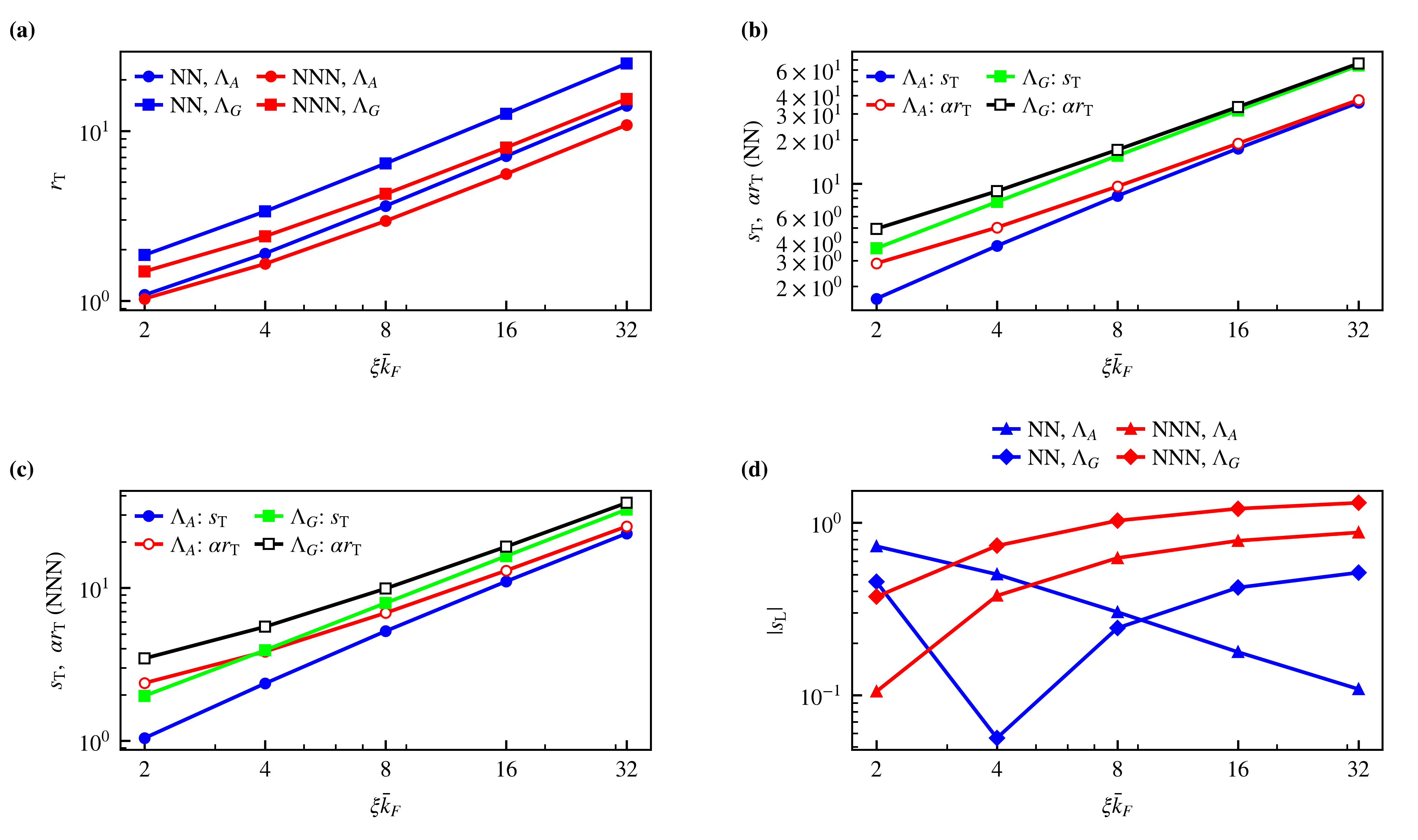}
\caption{Dependence on the correlation length at fixed density and $k_BT/t_1=0.05$. (a) Ratio $r_T$ of the spin flip quasiparticle scattering rate, averaged over the unsplit Fermi line, to $\Gamma_{T,\rm tr}$. In (b) NN and (c) NNN, filled circles and squares show $s_T$, and open circles and squares show $\alpha r_T$, with $\alpha=2.65$ on NN and $2.32$ on NNN from band integrals. (d) Filled triangles and diamonds show $|s_L|$ for $\Lambda_A$ and $\Lambda_G$, respectively.}\label{fig:l_xi}
\end{figure*}

\emph{Forward spin flips relax the spin current.}
A nearly forward flip turns the velocity only slightly, so it degrades the charge current only through the factor $1-\mathbf v'\cdot\mathbf v/|\mathbf v|^2$. The field drives each spin with its own velocity, so at equal energy the two spin populations differ by $F(u_x-d_{\mathbf k}\partial_Ev_x)$ per unit $\sigma\phi_0$, the velocity difference of the two spins at that energy. A forward flip moves an electron to the other spin band at the same energy, so it removes this imbalance at twice the quasiparticle scattering rate $\Gamma_{\rm out}$. The energy shift of the occupation escapes this relaxation. A flipped electron lands on the other spin band where the band shift had already placed the occupation, so $\Lcal_1$ cancels the spin flip relaxation of this part. We show this in the End Matter. Hence $s_T\simeq\alpha r_T+O(1)$, where $r_T=\Gamma^{\rm avg}_{T,\rm out}/\Gamma_{T,\rm tr}$, with $\Gamma_{T,\rm tr}$ twice the spin flip charge decay rate on the unsplit Fermi line, as for $K$, and band integrals give $\alpha=2.65$ on NN and $2.32$ on NNN. For Ornstein--Zernike fluctuations, the quasiparticle rate grows as $\xi$, because the integral over forward scattering along a two dimensional Fermi line grows as $\pi\xi$, while the momentum relaxation factor removes this growth from the charge rate. Hence $s_T$ grows nearly linearly with $\xi$ [Fig.~\ref{fig:l_xi} and Supplemental Material Sec.~S1]. On a three dimensional Fermi surface the two rates still separate, but only logarithmically (End Matter). On the lattice bands the spin conserving slope stays bounded, $|s_L|\le1.31$ for $2\le\xi\bar k_F\le32$, because for spin conserving scattering $\Lcal_-=\Lcal_+$ and $\Lcal_1$ carries $d_{\mathbf k}-d_{\mathbf k'}$, so $\Gamma_s$ and $\Gamma_c$ are both transport weighted rates.

\begin{figure}[tbp]
\centering\includegraphics[width=3.4in]{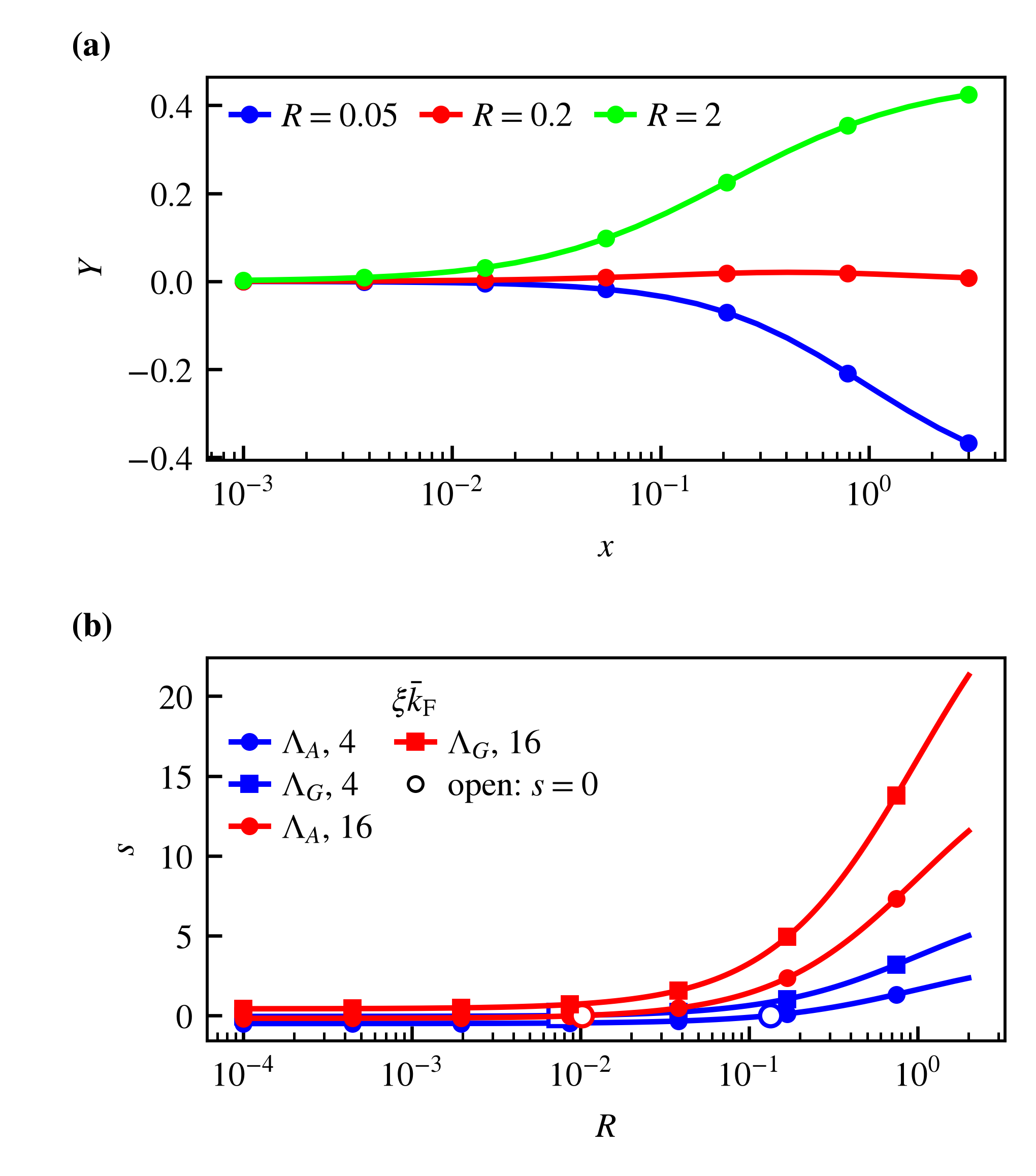}
\caption{Equal charge transport with different spin conversion on NN at fixed density and $k_BT/t_1=0.05$. (a) Relative reduction $Y=1-Q/Q_{\rm CRT}$ against relative magnetic resistivity $x$, for $\Lambda_A$ and $\xi\bar k_F=4$. At each total strength $A_L+A_T$, the three mixtures have the same charge response. (b) Initial slope $s$ against $R$. The enlarged open markers show the zeros at positive $R$. At $\xi\bar k_F=16$, both $s_L$ and $s_T$ of $\Lambda_G$ are positive, so its slope is positive at every $R$.}\label{fig:l_parts}
\end{figure}

The magnetic anisotropy decides how $\theta$ responds to critical fluctuations. Under easy axis anisotropy, the critical fluctuations are longitudinal, $R$ is small and $s\simeq s_L$, so the initial spin decay rate stays within a factor of order one of the charge decay rate at every $\xi$. For isotropic or soft transverse fluctuations, $s$ grows with $\xi$ while flips reach the other spin band. On NN, the amplitude ratio at which $s$ changes sign is $R^*=-s_L/s_T$, equal to $0.133$ for $\Lambda_A$ at $\xi\bar k_F=4$ and $0.0102$ at $16$ [Fig.~\ref{fig:l_parts}(b) and Supplemental Material Sec.~S2]. A rising $\theta$ requires almost purely longitudinal fluctuations, with a flip fraction below $R^*/(1+R^*)$.

Collisions that conserve the unsplit energy, $\Lcal_1=0$, give the wrong sign of $s$ on the NN band with $\Lambda_A$, $-1.57$ and $-1.95$ at $R=1$ and 2 against the full $1.64$ and $2.35$, and almost the whole difference comes from spin flip scattering (Supplemental Material Sec.~S3).

The lifetime approximation relaxes each state with its rate of scattering out, and the transport time approximation weights each transition by $1-\mathbf v'\cdot\mathbf v/|\mathbf v|^2$. Both replace scattering into a state by a local rate. By symmetry, the transport time approximation reproduces exactly the initial growth of the resistivity with magnetic scattering (End Matter), but it gives only 23 to 47 percent of $s$ on the reference bands, and the lifetime slope is at most $0.35$ (Supplemental Material Sec.~S4).

\emph{Parabolic band.}
In the parabolic band, $u_x=bv_x$ with $b=d_0/\mu_0$. Each spin band is then an anisotropic parabola with $1/m_{x\sigma}=(1+\sigma\phi_0b)/m$, $1/m_{y\sigma}=(1-\sigma\phi_0b)/m$ and the same density, so a constant relaxation time gives the Drude ratio $\theta=\phi_0b$ at every $\phi_0$, that is, $Q_{\rm CRT}=b$. Every deviation of $Q$ from $b$ comes from scattering. At $K=0.1$ and $a=3$, the full calculation gives $Q=0.675$ for $\Lambda_A$, whereas collisions on the unsplit bands give $1.34$. Energy conservation on the split bands thus reverses the sign of the correction.

\begin{figure}[tbp]
\centering\includegraphics[width=3.4in]{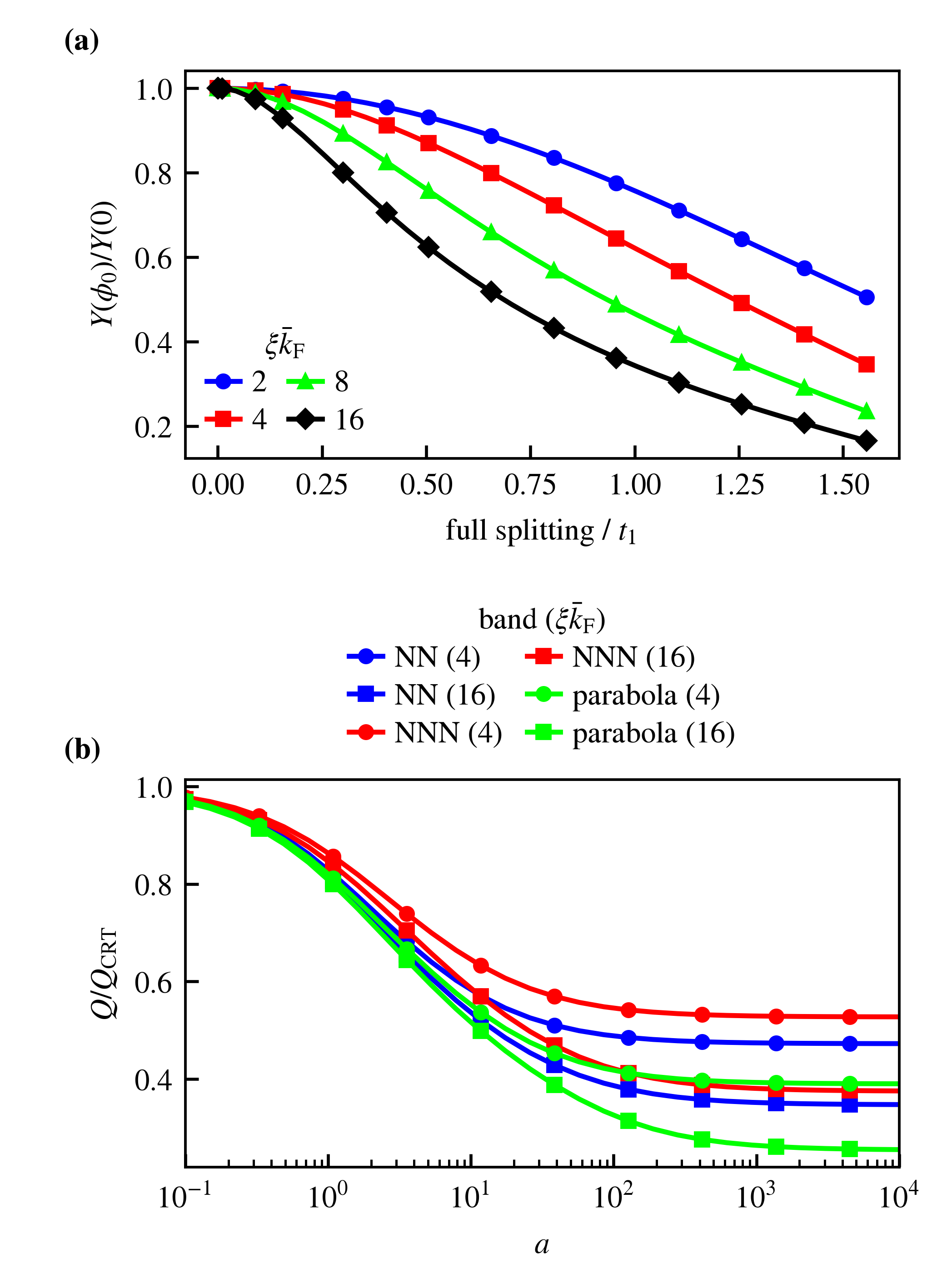}
\caption{Finite splitting and strong spin flips at fixed density. (a) Reduction $Y(\phi_0)=1-\theta(\phi_0)/\theta_{\rm bg}(\phi_0)$ divided by its value at $\phi_0\to0$, on NN with $\Lambda_A$, $R=2$ and total strength $K(1+R)=0.09$. The horizontal axis shows the full splitting $3\phi_0t_1$ of the Fermi line in units of $t_1$, up to $\phi_0=0.519$ (End Matter). (b) Full collision result for $Q/Q_{\rm CRT}$ at linear order against $a$, for $\Lambda_A$ and $K=0.01$. The parabola uses $b=1$.}\label{fig:latticecurrent}
\end{figure}

\emph{The splitting protects the spin current.}
A flip between the split Fermi lines needs a momentum transfer of at least $2\phi_0|d_{\mathbf k}|/(\hbar|\mathbf v|)$. Once this transfer exceeds $1/\xi$, the splitting itself blocks the forward flips. Equivalently, while an electron crosses a correlated region, its spin precesses by more than a radian in the exchange field, so it follows the slowly varying fluctuation axis adiabatically (End Matter). At $R=2$ and the full splitting $3\phi_0t_1=1.5t_1$ [Fig.~\ref{fig:latticecurrent}(a)], the reduction of $\theta$ is $0.53$, $0.37$, $0.26$ and $0.18$ of its linear value for $\xi\bar k_F=2$, $4$, $8$ and $16$, so the blocking becomes stronger with correlation length. For this NN bath, the linear response holds within one percent for $\xi\bar k_F\Delta_s/E_F=\phi_0\xi\bar k_F\lesssim0.5$ at long correlation length, with $\Delta_s=3\phi_0t_1$ the largest splitting on the Fermi line and $E_F=3t_1$. Because $\beta<\nu$ in the three dimensional Ising, XY and Heisenberg classes \cite{PelissettoVicari2002}, $\xi\Delta_s$ grows on approaching $T_N$ from below, so the splitting blocks the forward flips close to the transition, while further below they dominate in a window that widens as $\Delta_s/E_F$ decreases (End Matter).

\emph{Strong spin flips.}
Below $T_N$, soft transverse modes can make spin flips dominate, $R\gg1$. Once spin flips limit both currents, each decreases approximately as $1/a$ and their ratio settles on a plateau [Fig.~\ref{fig:latticecurrent}(b), where $R=f_ma/K$ exceeds $10^2$], independent of $K$. The occupation that follows the band shift decays only with the charge current, and forward flips equalize the two spin distributions at equal energy, so the plateau approaches the spin current of the charge carrying distribution placed on the two split Fermi lines (End Matter). For $\Lambda_A$, the plateau retains $0.47$ of the band ratio on NN and $0.53$ on NNN at $\xi\bar k_F=4$, and $0.35$ and $0.37$ at $16$. Reaching the plateau requires clean samples, with $E_F\tau_{\rm bg}/\hbar\gtrsim4\times10^2$ at $\xi\bar k_F=4$ and up to about $10^4$ at $16$.

\emph{Disorder series.}
Samples that differ only in background disorder share $\phi_0$ and the fluctuation strengths, so changing $\tau_{\rm bg}$ moves a sample along a line of fixed $R$. With a constant background, $\theta=\phi_0Q_{\rm CRT}$, so $\theta/\theta_{\rm bg}=Q/Q_{\rm CRT}$, where $\theta_{\rm bg}$ is the spin splitter ratio of samples dominated by background scattering, extrapolated within the metallic regime. The measured fall $1-\theta/\theta_{\rm bg}$ against $\rho_{\rm mag}/\rho_{\rm bg}$ then gives $s$ directly, independent of $\phi_0$. Here $\rho_{\rm bg}$ is the residual resistivity and $\rho_{\rm mag}$ the magnetic resistivity, obtained for example by subtracting the resistivity of an isostructural nonmagnetic reference, and phonons enter through the shift below. To leading order, $\rho_{\rm mag}$ is independent of $\tau_{\rm bg}$, as in the scaling analysis that separates residual and phonon contributions to the anomalous Hall effect \cite{Tian2009}. On the reference bands at $\xi\bar k_F=4$, the full slope is $1.4$ to $2.4$ for $\Lambda_A$ and up to $5.0$ for $\Lambda_G$. Because $s_L$ stays of order one while $s_T$ grows in proportion to $\xi$, a slope that grows in proportion to $\xi$ toward $T_N$ signals spin flip critical fluctuations independent of band details, and its size constrains $R$ given $s_L$ and $s_T$ from a band calculation.

The measured slope equals $s$ for samples with the same ordered moment and magnetic spectrum, compared at equal $T/T_N$, and for nonmagnetic defects that leave the magnetic fluctuations unchanged. Anisotropic phonon scattering changes $\theta$ by itself \cite{He2025} and shifts the measured slope by $s_{\rm ph}\rho_{\rm ph}/\rho_{\rm mag}$ at every disorder level, a shift that stays small for $\rho_{\rm ph}$ below about $\rho_{\rm mag}$ \cite{JiangCompanion2026}. In a torque measurement, a thickness series at each disorder level fixes the disorder dependence of the interface and spin diffusion factor. Layered KV$_2$Se$_2$O \cite{Jiang2025Metallic}, closest to the two dimensional kinematics treated here, and Mn$_5$Si$_3$ films \cite{Reichlova2024} are possible test beds.

Critical magnetic fluctuations thus treat charge and spin currents differently. Spin conserving fluctuations keep the initial spin decay rate of order the charge decay rate, whereas nearly forward spin flips equalize the two spin populations at equal energy at a quasiparticle rate that grows with the correlation length. A disorder series at fixed reduced temperature measures this response independent of the order amplitude, and its growth toward $T_N$ identifies spin flip critical fluctuations.

\FloatBarrier

\section*{End Matter}
\subsection*{Scattering strengths}
The periodic lattice momentum transfer is $q^2=4\sum_\alpha\sin^2[(k_\alpha-k'_\alpha)/2]$, and both couplings are periodic in each momentum separately. In the parabolic band, $q=|\mathbf k-\mathbf k'|$ in units $m=\mu_0=\hbar=1$. The radius $\bar k_F=\oint d\ell\,|\mathbf k|/|\mathbf v|\big/\oint d\ell/|\mathbf v|$ is weighted by the density of states, and on the NN band $\xi\bar k_F=4$ gives $\xi=2.02$ lattice spacings. The strength $A_T$ sums both transverse Cartesian components. The spin conserving strength $K=2\tau_{\rm bg}\langle v_x,L_Lv_x\rangle_w/\langle v_x,v_x\rangle_w$ is twice the longitudinal charge decay rate on the unsplit Fermi line, in units of $1/\tau_{\rm bg}$, and the spin flip strength $a=2\tau_{\rm bg}\Gamma^{\rm AN}_{T,\rm out}$ is twice the rate of transverse scattering out of the antinode, the point of largest $|\Phi_d|$. Because a flip changes the spin imbalance by two, $a$ is also $\tau_{\rm bg}$ times the relaxation rate of a uniform spin imbalance at the antinode. Both strengths are calibrated once at zero splitting. They measure different properties of the scattering, so $R=f_ma/K$, with $f_m=0.150,0.0819$ on the NN band and $0.175,0.116$ on the NNN band for $\Lambda_A,\Lambda_G$. At $R=2$, $a/K=2/f_m$ grows nearly linearly with $\xi$, because $f_m$ falls as the quasiparticle scattering rate grows (Supplemental Material Sec.~S1). A chiral transverse correlator has the symmetry of the response computed here, because altermagnetic order splits magnons of opposite chirality with a $d$-wave dependence. It drops out only because it is smaller by $\hbar\omega_q/k_BT$, which is small near $T_N$, where the fluctuations slow down.

\subsection*{Background and fixed density}
Pointlike impurities that conserve spin give $\tau_{\rm bg}^{-1}(E)\propto N(E)$, equal for both spins at equal energy, and their scattering into a state vanishes on the odd distributions driven by the field. At fixed momentum, the drive gains $d_{\mathbf k}\tau'_{\rm bg}Fv_x$, which combines with the energy shift into the occupation contribution $-\hbar^{-1}\int_{\mathbf k}d_{\mathbf k}\tau_{\rm bg}F\partial_{k_x}v_x$ after integration by parts. The ratio changes only through thermal averaging. At $k_BT/t_1=0.05$, the NN and NNN reference ratios shift by $-0.12$ and $-0.46$ percent, and the slopes change by less than 1 percent (Supplemental Material Sec.~S5).

Density, temperature and fluctuation strengths stay fixed as $\phi_0$ varies, which is exact at linear order because the strengths are even in $\phi_0$. We solve $n(\mu(\phi_0,T),\phi_0,T)=n(\mu_0,0,0)$, where $n$ counts both spins, so $\partial\mu/\partial\phi_0=0$ at zero splitting. The condition $\hbar\Gamma^{\rm tot}_{\rm out}/E_F<0.1$ requires $E_F\tau_{\rm bg}/\hbar>10\tau_{\rm bg}\Gamma^{\rm tot}_{\rm out}$, with $E_F=\mu-\epsilon_{\min}$. For the NN band with $\Lambda_A$, $K=0.1$ and $R=2$, this bound at the antinode is about 20. The Born approximation for fluctuations correlated over $\xi$ also requires a mean free path $|\mathbf v|/\Gamma_{\rm out}$ longer than $\xi$, because the Born approximation treats the phase that an electron accumulates while crossing a correlated region as small, and its square is of order $\xi\Gamma_{\rm out}/|\mathbf v|$.

\subsection*{Collision derivative and relaxation time approximations}
For a distribution $h$ that includes $F$,
\begin{align}
(\Lcal_1h)_{\mathbf k}&=\int_{\mathbf k'}[\mathcal A_L(d_{\mathbf k}-d_{\mathbf k'})+\mathcal A_T(d_{\mathbf k}+d_{\mathbf k'})]\nonumber\\
&\quad\times\delta'(\epsilon_0(\mathbf k)-\epsilon_0(\mathbf k'))(h_{\mathbf k}-h_{\mathbf k'}),
\label{eq:l_m1}
\end{align}
where $\mathcal A_{L,T}=A_{L,T}|\Lambda_m|^2/(\xi^{-2}+q^2)$. The terms with $h_{\mathbf k}$ and $-h_{\mathbf k'}$ describe scattering out of and into $\mathbf k$, and integration over the final energy turns $\delta'$ into a positive energy derivative of the integral over the final energy shell. $\Lcal_1$ describes how the order changes which final states the scattering can reach.

The lifetime approximation uses $\tau_\sigma=\tau_{\rm bg}/(1+\tau_{\rm bg}\Gamma_{{\rm out},\sigma})$. The transport time approximation replaces this rate by $\sum_{\mathbf k'\sigma'}W[1-\mathbf v'\cdot\mathbf v/|\mathbf v|^2]$, where $W$ is the transition rate. To first order, the full charge loss is $\tau_{\rm bg}\sum FW(v_x^2-v_xv'_x)$ and the transport time loss is $\tau_{\rm bg}\sum FWv_x^2(1-\mathbf v\cdot\mathbf v'/|\mathbf v|^2)$. Elasticity gives $F_{\mathbf k}=F_{\mathbf k'}$, so $FW$ is symmetric. Invariance under $k_x\leftrightarrow k_y$ gives $\sum FWv_x^2\mathbf v\cdot\mathbf v'/|\mathbf v|^2=\tfrac12\sum FW\mathbf v\cdot\mathbf v'=\sum FWv_xv'_x$, so the two charge slopes are equal. The lifetime charge slopes are larger by factors of 2.2 to 5.4.

\subsection*{Forward spin flips}
Elasticity combines the occupation source and the collision derivative into
\begin{equation}
\Lcal_-X_1+\Lcal_1X_0=F(\Lcal_-u_x+\Lcal_1v_x)+F'd_{\mathbf k}\Lcal_+v_x,
\label{eq:l_a1}
\end{equation}
where $\Lcal_1v_x$ is the operator of Eq.~(\ref{eq:l_m1}) acting on the bare $v_x$. The same identity holds with $v_x$ replaced by any distribution $g$. With $h_+=Fg$ and $g=(I+\tau_{\rm bg}\Lcal_+)^{-1}v_x$, it gives the exact solution of Eq.~(\ref{eq:l_kinetic}),
\begin{equation}
h_-=d_{\mathbf k}F'g+F(I+\tau_{\rm bg}\Lcal_-)^{-1}(u_x-\tau_{\rm bg}\Lcal_1g).
\label{eq:l_exact}
\end{equation}
The part of the occupation that only shifts the band energy is $d_{\mathbf k}F'g$ at every scattering strength, and it
relaxes through momentum relaxation alone, because a flipped electron lands where the band shift had already placed the occupation. In the forward limit, $\Lcal_{+,T}v_x\to0$, $\Lcal_{-,T}\to2\Gamma_T$ and $\Lcal_{1,T}g\to-2d_{\mathbf k}\Gamma_T\partial_Eg$, with $\partial_E=(\mathbf v\cdot\nabla_{\mathbf k})/(\hbar|\mathbf v|^2)$. Here $u_x-d_{\mathbf k}\partial_Ev_x$ is the change of $v_x$ per unit $\sigma\phi_0$ at fixed energy, because the spin $\sigma$ line at energy $E$ lies a distance $-\sigma\phi_0d_{\mathbf k}/(\hbar|\mathbf v|)$ from the unsplit line along its normal, so with background scattering alone $F(u_x-d_{\mathbf k}\partial_Ev_x)=h_--d_{\mathbf k}\partial_Eh_+$ is the imbalance of the two spin populations at equal energy. At weak scattering, forward flips remove it at the rate $2\Gamma_T$, and the remaining spin rate is $2\int_{\mathbf k}F\Gamma_Tv_x(u_x-d_{\mathbf k}\partial_Ev_x)/S_0$. For Ornstein--Zernike fluctuations, $\Gamma_T\propto\Phi_d^2/|\mathbf v|$, which gives the coefficient $\alpha=2.65$ on the NN band and $2.32$ on the NNN band. The integral of $(\xi^{-2}+q^2)^{-1}$ along the Fermi line grows as $\pi\xi$, while the velocity difference in the transport integral removes this growth, so $r_T$ grows as $\xi$. On a three dimensional Fermi surface, $\int d^2q\,(\xi^{-2}+q^2)^{-1}\simeq2\pi\ln(\xi q_c)$, with $q_c$ set by the band, while the transport integral stays finite, so $r_T\propto\ln(\xi\bar k_F)$. The fitted exponents and the thermal correction to $\alpha r_T$ are given in Supplemental Material Sec.~S1.

When flips limit both currents, $\tau_{\rm bg}\Lcal_\pm\gg1$, Eq.~(\ref{eq:l_exact}) makes $S$ and $C$ both scale as $1/a$, so their ratio depends only on the bands and on the shape of the scattering. In the forward limit, the last term of Eq.~(\ref{eq:l_exact}) then approaches $Fd_{\mathbf k}\partial_Eg$ up to corrections of order $1/r_T$, and
\begin{equation}
S\to\int_{\mathbf k}\left[u_xFg+v_xd_{\mathbf k}\partial_E(Fg)\right],
\label{eq:l_plateau}
\end{equation}
the spin current of a common distribution $Fg$ placed at equal energy on the two split bands.

\subsection*{Finite splitting}
We evaluate $\theta(\phi_0)$ and its background at the same density and splitting, keeping the fluctuation strengths fixed, and stop at $\phi_0=0.519$, where the first spin saddle lies $3k_BT$ from the chemical potential. A forward flip requires $2\phi_0|d_{\mathbf k}|\xi/(\hbar|\mathbf v|)\ll1$, and Fig.~\ref{fig:latticecurrent}(a) follows the response beyond this window. The blocking is lifted only near the nodes, where both the splitting and the $d$-wave coupling vanish. It also requires that fluctuations at this momentum transfer carry an energy small compared with the splitting, which for magnons means a magnon velocity well below $|\mathbf v|$. Below $T_N$, $\xi\Delta_s\propto(1-T/T_N)^{\beta-\nu}$ grows only slowly, with $\beta-\nu\simeq-0.3$. For a rough estimate we take Heisenberg exponents, $\Delta_s=\Delta_s(0)(1-T/T_N)^\beta$, $\xi=\xi_0(1-T/T_N)^{-\nu}$ and $\xi_0=1/\bar k_F$. The condition $\xi\bar k_F\Delta_s/E_F\lesssim0.5$ then holds up to $\xi\bar k_F\approx30$ for $\Delta_s(0)=0.1E_F$, but only up to $\xi\bar k_F\approx3$ for $\Delta_s(0)=0.3E_F$. An injected spin current additionally requires boundary conditions and spatial kinetics \cite{Sigales2026,Zarzuela2025}.

\clearpage
\setcounter{section}{0}
\setcounter{subsection}{0}
\setcounter{figure}{0}
\setcounter{table}{0}
\setcounter{equation}{0}

\makeatletter
\def\@hangfrom@section#1#2#3{\@hangfrom{#1#2}\MakeTextUppercase{#3}}
\let\@sectioncntformat\@seccntformat
\makeatother
\counterwithout*{equation}{section}
\setcounter{secnumdepth}{1}
\renewcommand{\thesection}{S\arabic{section}}
\renewcommand{\thesubsection}{S\arabic{section}.\arabic{subsection}}
\renewcommand{\thefigure}{S\arabic{figure}}
\renewcommand{\thetable}{S\arabic{table}}
\renewcommand{\theequation}{S\arabic{equation}}

\providecommand{\theHsection}{}
\renewcommand{\theHsection}{SM.\arabic{section}}
\providecommand{\theHsubsection}{}
\renewcommand{\theHsubsection}{SM.\arabic{section}.\arabic{subsection}}
\providecommand{\theHfigure}{}
\renewcommand{\theHfigure}{SM.\arabic{figure}}
\providecommand{\theHtable}{}
\renewcommand{\theHtable}{SM.\arabic{table}}
\providecommand{\theHequation}{}
\renewcommand{\theHequation}{SM.\arabic{equation}}
\onecolumngrid
\begin{center}
{\large\bfseries Supplemental Material for ``Spin conservation protects the altermagnetic spin-splitter current from critical scattering''\par}
\vspace{0.8em}
Yunxi Jiang and Hao Yu
\end{center}

\twocolumngrid

\renewcommand{\citenumfont}[1]{#1}

NN denotes the band with only nearest neighbor hopping and NNN the band that also includes next nearest neighbor hopping. $K$ and $a$ are the dimensionless spin conserving and spin flip strengths defined in the End Matter of the Letter, and $R=A_T/A_L$ is the ratio of transverse to longitudinal fluctuation strengths.

\section{Correlation length}
The rates are calibrated on the unsplit Fermi line at each $\xi$ and then held fixed as the order amplitude varies. For Ornstein--Zernike fluctuations, the integral for scattering out of a state grows with $\xi$, while the transport weight suppresses the small momentum transfers responsible for this growth. The transport rate approaches a finite value, so $r_T$ and $s_T$ grow nearly linearly. Fits over $\xi\bar k_F=8,16,32$ give exponents $0.93$ to $0.98$ for $r_T$ and $1.02$ to $1.06$ for $s_T$. At $\xi\bar k_F=32$, the ratios $s_T/(\alpha r_T)$ are $0.957$ and $0.968$ on NN and $0.899$ and $0.906$ on NNN, for $\Lambda_A$ and $\Lambda_G$. The remaining difference from unity comes from the thermal average of the charge rate and from the finite terms in $s_T\simeq\alpha r_T+O(1)$. Over $2\le\xi\bar k_F\le32$, $|s_L|$ is at most $1.31$, because a forward spin conserving transition carries the weight $d_{\mathbf k}-d_{\mathbf k'}\simeq0$.

At $R=2$, $a/K=2/f_m$. For $\Lambda_A$ this ratio is $13.35$, $27.51$, $57.05$ and $116.96$ on NN and $11.41$, $22.76$, $47.26$ and $97.97$ on NNN at $\xi\bar k_F=4$, $8$, $16$ and $32$. For $\Lambda_G$ it is $24.42$, $49.93$, $102.43$ and $208.42$ on NN and $17.17$, $33.61$, $68.60$ and $140.60$ on NNN. The falling calibration factor $f_m$ turns a bounded longitudinal transport strength into a growing transverse quasiparticle scattering rate.

\section{Sign threshold and correlation length}
A positive threshold $R^*=-s_L/s_T$ exists when $s_L$ and $s_T$ have opposite signs. On NN, $\Lambda_A$ has thresholds $0.444$, $0.133$, $0.0366$, $0.0102$ and $0.00303$ at $\xi\bar k_F=2$, $4$, $8$, $16$ and $32$. For $\Lambda_G$ they are $0.125$ and $0.00747$ at $2$ and $4$. At $8$, $16$ and $32$ both channel slopes of $\Lambda_G$ are positive, so $s$ is positive at every $R$. Figure~2(b) of the Letter shows the change from $4$ to $16$.

\section{Spin conserving and spin flip limits and the collision derivative}
At $k_BT/t_1=0.05$, the spin conserving and spin flip limits of the slope are $s_L=-0.50$ and $s_T=3.77$ on the NN band, $0.38$ and $2.37$ on the NNN band, and $0.20$ and $8.41$ in the parabolic band, for $\Lambda_A$. For $\Lambda_G$ they are $-0.056$ and $7.53$, $0.74$ and $3.92$, and $2.40$ and $20.29$. On the NN band $s_L<0<s_T$, which gives $R^*=-s_L/s_T=0.133$ for $\Lambda_A$ and $0.0075$ for $\Lambda_G$. Setting $\mathcal L_1=0$ removes the part of the slope that comes from the change of the collisions with the order. For $\Lambda_A$ the spin conserving part of this missing slope is $-0.065$ on the NN band and $0.055$ on the NNN band, and its spin flip part is $6.47$ and $1.81$. The spin conserving part stays below $0.11$ on both lattice bands for both couplings, because $d_{\mathbf k}-d_{\mathbf k'}$ vanishes for forward scattering. Almost all of the missing slope therefore comes from spin flip scattering. The companion Article \cite{SMJiangCompanion2026} resolves the slope into velocity and occupation contributions.

\section{Disorder series}
\begin{figure}[tbp]\centering\includegraphics[width=3.4in]{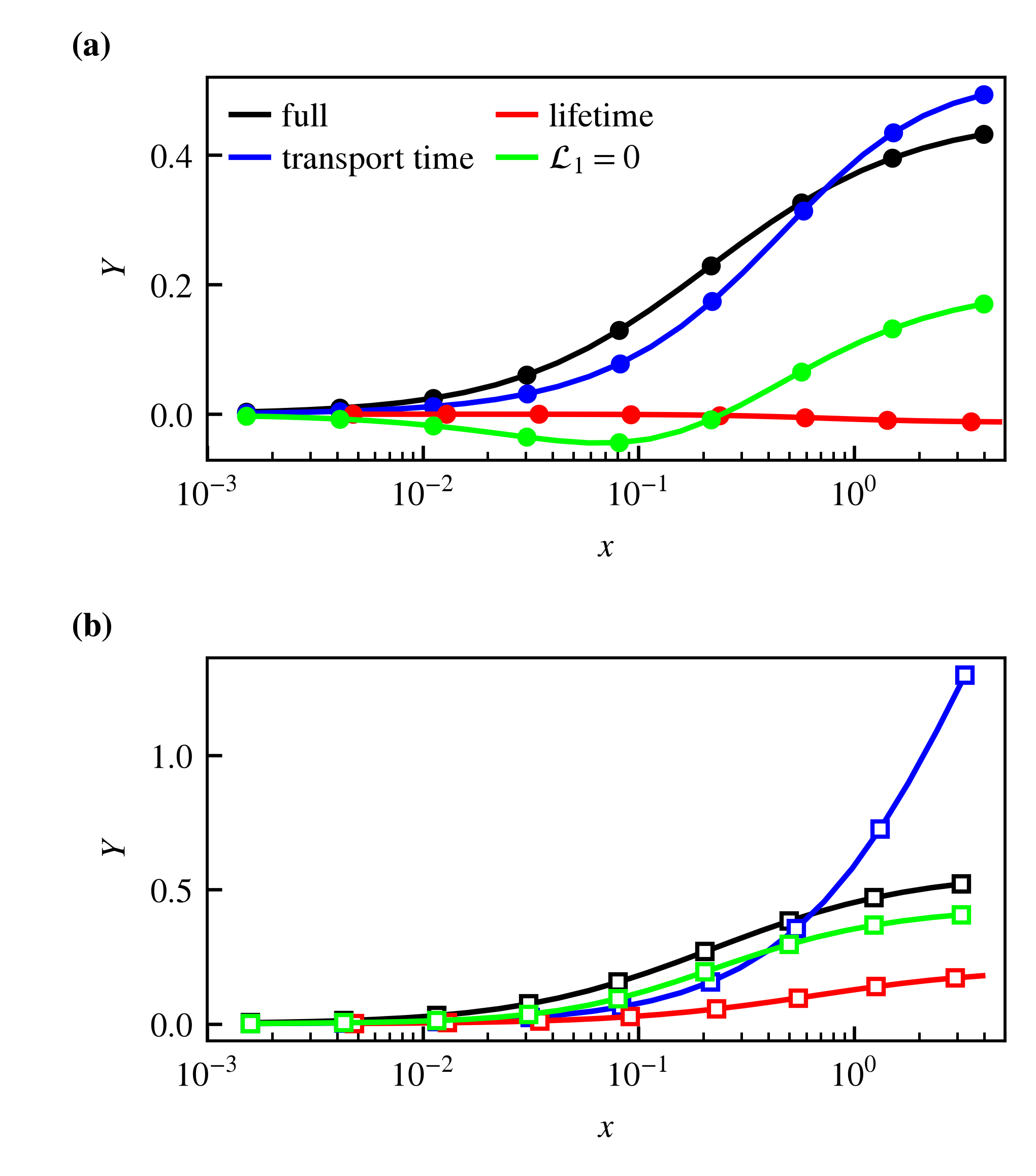}
\caption{Disorder series at $R=2$ for (a) NN with $\Lambda_A$ and (b) NNN with $\Lambda_G$, at fixed density, $k_BT/t_1=0.05$ and $\xi\bar k_F=4$. Black denotes full collisions, blue transport time, red lifetime and green unsplit collisions with $\mathcal L_1=0$. Each model uses its own charge response, $x=C_0/C-1$, against $Y=1-Q/Q_{\rm CRT}$. Negative $Y$ means enhancement of the reduced spin splitter ratio.}\label{fig:sm_series}\end{figure}

Figure~\ref{fig:sm_series} shows two disorder series at $R=2$, NN with $\Lambda_A$, where the unsplit collision model gives the wrong initial sign and changes its trend at stronger scattering, and NNN with $\Lambda_G$, where the transport time approximation strongly overestimates the reduction at larger resistivity. Table~\ref{tab:sm_series} summarizes the initial slopes for all seven band, coupling and bath combinations. Fits of $Y=sx/(1+cx)$ on $0<x<1$, with $c$ describing the curvature, give a slope up to 10 percent below its value at the origin. A measured series is therefore best compared with the full calculated curve, or fitted in the range $sx\lesssim0.04$, in which the calculated series stay within 10 percent of the linear law \cite{SMJiangCompanion2026}.

\begin{table}[tbp]
\caption{Initial slopes $s=\lim_{x\to0}Y/x$ from the first order rates for the seven band, coupling and bath combinations of the disorder series. TT and LT denote transport time and lifetime, and the last column uses collisions on the unsplit bands. Each approximation uses its own $x=C_0/C-1$. The temperature and correlation length are those of Fig.~\ref{fig:sm_series}, and $A$ and $G$ denote $\Lambda_A$ and $\Lambda_G$.}\label{tab:sm_series}
\begin{ruledtabular}
\begin{tabular}{lrrrrrr}
Band & Coupling & $R$ & Full & TT & LT & $\mathcal L_1=0$\\
\hline
NN & $A$ & 1 & 1.64 & 0.52 & -0.26 & -1.57\\
NN & $G$ & 1 & 3.74 & 1.43 & -0.26 & -2.47\\
NNN & $A$ & 1 & 1.38 & 0.33 & 0.18 & 0.45\\
NNN & $G$ & 1 & 2.33 & 0.54 & 0.21 & 0.83\\
NN & $A$ & 2 & 2.35 & 1.08 & -0.001 & -1.95\\
NN & $G$ & 2 & 5.00 & 2.37 & 0.02 & -3.26\\
NNN & $G$ & 2 & 2.86 & 0.81 & 0.35 & 0.90\\
\end{tabular}
\end{ruledtabular}
\end{table}

\section{Pointlike impurities}
For an elastic pointlike impurity that conserves spin, a constant scattering amplitude gives $\Gamma_{\rm imp}(E)\propto N(E)$. Inversion makes its scattering into a state vanish on the odd distributions driven by the field, and opposite spins have equal densities of states at equal energy. We normalize to $\tau_{\rm bg}\Gamma_{\rm imp}=1$ at the unsplit Fermi level and keep the amplitude fixed at all energies and order amplitudes. At $k_BT/t_1=0.05$, the reference ratio changes by $-0.12$ percent on the NN band and $-0.46$ percent on the NNN band. Near the van Hove saddle, $N(E)$ varies sharply inside the thermal window. For $(t_2/t_1,\nu,k_BT/t_1)=(-0.2,0.40,0.02)$ the chemical potential lies $2.09k_BT$ from the saddle and the ratio shifts by $-2.4$ percent, and at $k_BT/t_1=0.05$ the same density lies $0.94k_BT$ from the saddle and the shift is $-3.8$ percent. With this background the slopes change by at most $0.98$ percent on the lattice bands, for the NNN band with $\Lambda_A$ at $R=1$, and stay the same in the parabolic band. The transport time approximation still gives a quarter to a half of the full slope.

\FloatBarrier

\makeatletter
\let\thebibliography\NAT@thebibliography
\let\endthebibliography\endNAT@thebibliography
\def\place@bibnumber#1{[#1]}
\makeatother

\end{document}